\documentclass[11pt,a4paper]{article}
\usepackage{jheppub}

\usepackage{afterpage}

\usepackage{amssymb}
\usepackage{amsmath}
\usepackage{graphicx}
\usepackage{longtable}
\usepackage{verbatim}
\usepackage{amsfonts}
\usepackage{xcolor}

\newcommand{\be}{\begin{equation}}
\newcommand{\ee}{\end{equation}}
\newcommand{\beq}{\begin{equation}}
\newcommand{\eeq}{\end{equation}}

\newcommand{\iu}{{\mathrm i}}
\newcommand{\E}{{\mathrm e}}

\newcommand{\tF}{{\widetilde F}}

\newcommand{\tR}{{\widetilde R}}
\newcommand{\Mpl}{{M_{\rm pl}}}

\usepackage{xcolor}
\usepackage{bm}
\usepackage[normalem]{ulem}

\vspace*{-30mm}

\title{\boldmath Clockwork Axions in Cosmology\\
{\large Is Chromonatural Inflation Chrononatural?}}
\author[a]{Prateek Agrawal,}
\author[b]{JiJi Fan,} 
\author[a]{Matthew Reece}
\affiliation[a]{Department of Physics, Harvard University, Cambridge, MA 02138, USA}
\affiliation[b]{Department of Physics, Brown University, Providence, RI, 02912, USA}

\vspace*{1cm}

\abstract{ Many cosmological models rely on large couplings of axions
  to gauge fields. Examples include theories of magnetogenesis,
  inflation on a steep potential,  chiral gravitational waves,  and
  chromonatural inflation. Such theories require a mismatch between
  the axion field range and the mass scale appearing in the $a F
  \widetilde{F}$ coupling. This mismatch suggests an underlying
  monodromy, with the axion winding around its fundamental period a large
  number of times. We investigate the
  extent to which this integer can be explained as a product of
  smaller integers in a UV completion: in the parlance of our times,
  can the theory be ``clockworked''? We argue that a clockwork
  construction producing a potential $\mu^4 \cos(\frac{a}{j F_a})$ for
  an axion of fundamental period $F_a$ will obey the constraint $\mu <
  F_a$. For some applications, including chromonatural inflation with
  sub-Planckian field range, this constraint obstructs a clockwork UV
  completion.  Alternative routes to a large coupling include fields
  of large charge (an approach limited by strong coupling) or kinetic
  mixing (requiring a lighter axion). Our results suggest that
  completions of axion cosmologies that explain the large parameter in
the theory potentially alter the phenomenological predictions of the
model.  }

\begin{document}

\maketitle

\section{Introduction: Large Couplings in Axion Cosmology}

A large set of cosmological models rely on a (pseudo)scalar field coupled to an $F \tF$ term for a gauge field or an $R \tR$ term for gravity. Applications include the generation of primordial magnetic fields \cite{Turner:1987bw, Garretson:1992vt,Adshead:2016iae}; dissipation allowing for inflation in steep potentials \cite{Anber:2009ua, Notari:2016npn, Ferreira:2017lnd}; baryo- or leptogenesis \cite{Alexander:2004us}; chromonatural inflation \cite{Adshead:2012kp,Dimastrogiovanni:2012st,Maleknejad:2012fw,Adshead:2013nka}; production of chiral gravitational waves during inflation \cite{Lue:1998mq,Alexander:2004wk,Anber:2012du,Adshead:2013qp,Maleknejad:2016qjz,Obata:2016xcr,Dimastrogiovanni:2016fuu}; preheating \cite{Adshead:2015pva}; decreasing the abundance of QCD axion dark matter \cite{Agrawal:2017eqm, Kitajima:2017peg} and providing alternative dissipation mechanisms \cite{Hook:2016mqo, Tangarife:2017vnd, Tangarife:2017rgl} in relaxion cosmology \cite{Graham:2015cka}.

Some of these theories are particularly interesting because they allow
qualitatively new phenomena compared to conventional theories. For
instance, the usual manner in which inflation can produce a possibly
observable tensor-to-scalar ratio is large-field inflation, that is,
if the inflaton field range is larger than the Planck scale. This is
the so-called ``Lyth bound'' \cite{Lyth:1996im}. Theories like
chromonatural inflation can produce a large signal from {\it chiral}
primordial gravitational waves, even with a small field range, thus
evading the Lyth bound. The central element is the
presence of gauge fields which are constantly replenished by the
rolling inflaton field. These gauge fields have tensor fluctuations
which are not related to scalar fluctuations in the same way as in
usual slow-roll inflation. Another interesting result is
that axion couplings to gauge fields, by providing a source of
dissipation other than Hubble friction, can allow inflation in steep
potentials that fail the usual slow-roll criteria \cite{Anber:2009ua}.

Both of these examples have heightened interest due to various
``Swampland'' conjectures about what is possible in quantum gravity
\cite{Vafa:2005ui, ArkaniHamed:2006dz, Ooguri:2006in}. For example,
theories of large-field inflation run into tension with the failure of
explicit searches for super-Planckian field ranges in string theory
\cite{Banks:2003sx,
Rudelius:2014wla,Conlon:2016aea,Baume:2016psm,McAllister:2016vzi,Cicoli:2018tcq,
Blumenhagen:2018nts}, with the Weak Gravity Conjecture
\cite{ArkaniHamed:2006dz,Rudelius:2015xta,Montero:2015ofa,Brown:2015iha,Bachlechner:2015qja,Hebecker:2015rya,Brown:2015lia,Heidenreich:2015wga},
and with more general Swampland arguments about difficulties with
super-Planckian field ranges
\cite{Ooguri:2006in,Nicolis:2008wh,Klaewer:2016kiy,Valenzuela:2016yny,Dolan:2017vmn,Blumenhagen:2017cxt,Grimm:2018ohb,Heidenreich:2018kpg,Landete:2018kqf}.
These difficulties make chromonatural inflation particularly
interesting as a theory that can generate a large tensor signal while
possibly evading such constraints. More recently, some authors have
speculated that there are no consistent de Sitter vacua in quantum
gravity
\cite{Brennan:2017rbf,Sethi:2017phn,Danielsson:2018ztv,Moritz:2018sui,Rajaraman:2016nvv}
(also see \cite{Kutasov:2015eba,Moritz:2017xto}). An even stronger
statement has been suggested: there is no region of the scalar
potential in quantum gravity for which $V > 0$ and $M_{\rm pl}
|\partial_\phi V|/V \ll 1$, putting tension on the slow-roll inflation
paradigm~\cite{Obied:2018sgi,swampland2:2018}.  This makes inflating
in steep potentials
through dissipation an even more interesting possibility, as it evades
such hypothetical (but speculative) Swampland bounds. For these
reasons, it is very interesting to consider the model-building of
these theories in more detail, to assess how plausible UV completions
may be.

A common feature for the cosmological models mentioned above is an axion-gauge field coupling that is parametrically large relative to the field range of the axion. 
An axion is a periodic field, $a \cong a + 2 \pi F_a$ with $F_a$ being the fundamental period.  
A typical Lagrangian in these models takes the form
\begin{align}
{\cal L} &\supset \mu^4\cos\left(\frac{a}{j F_a}\right)
+ k \frac{\alpha}{8 \pi} \frac{a}{F_a} F_{\mu \nu} \tF^{\mu \nu},
\label{eq:axion}
\end{align}
where the first term is the effective potential of the axion generated
(for example) by non-perturbative dynamics of some confining gauge
group with a confinement scale $\mu$. The
coefficient $j$ appears to violate this shift symmetry, but if it is
an integer it can arise via monodromy, i.e.~a potential with multiple
branches \cite{Witten:1980sp, Witten:1998uka, Silverstein:2008sg}.
In the second term, $F_{\mu\nu}$ is the field strength of some other
Abelian or non-Abelian gauge group, which will play different roles
depending on the model. The coefficient $k$ must be an integer due to the axion shift symmetry. 
The effective field range of the axion is then
\be
f_a \equiv j F_a.
\ee
Note that when gauge fields are canonically normalized, the
presence of the factor of $\alpha = \frac{g^2}{4\pi}$ with $g$ being
the gauge coupling is {\it required} for consistency with the discrete
shift symmetry of the axion. Unsurprisingly, the same factor appears
in the $\theta$-term for the gauge theory. Quite a few
phenomenological applications in the cosmological models simply write
the axion-gauge field couplings as $\frac{\lambda}{f_a} aF_{\mu \nu}
\tF^{\mu \nu}$ without explicitly writing the coupling
strength.\footnote{Be aware of notation: some literature used $\alpha$
instead of $\lambda$ to denote the overall axion-gauge field coupling,
but this $\alpha$ is not $g^2/(4\pi)$.} These applications typically
require the dimensionless coefficient $\lambda \gg 1$. In our
notation in Eq.~\ref{eq:axion}, this requirement is translated to 
\be
k j \alpha \gg 1. 
\label{eq:enhanced}
\ee 
This factor of $\alpha = g^2/(4\pi)$ has not been emphasized in
the literature of many cosmological models. In perturbative models,
$\alpha \lesssim 0.1$ when the gauge coupling is $\sim 1$. Sometimes
there are further constraints on the size of the gauge coupling and
$\alpha$ has to be smaller.  Taking $\alpha$ into account, an enhanced
axion-gauge field coupling usually requires a huge value for $k\times
j$, which could be orders of magnitude larger than the number quoted
in the literature.  While there are ways of getting enhanced couplings
of axions to gauge fields~\cite{Farina:2016tgd, Agrawal:2017cmd},
these mechanisms are subject to simple theoretical constraints. It is
thus highly non-trivial and interesting to investigate whether and how
to get the large axion-gauge field coupling needed for interesting
cosmological dynamics.

We see from the discussion above that there are two enhancement
factors, an enhanced coupling $k$ and an enhanced field range
$j$. 
There are three possible types of mechanisms to generate a large
coupling between the compact axion field and gauge fields. The first
two enhance the coupling via increasing $k$, while the third enhances
the field range.
\begin{itemize}
\item Inclusion of matter fields such as vector-like fermions with
  large charges under the gauge group or large PQ charge;
\item Kinetic mixing between multiple axions \cite{Babu:1994id};
\item Two axion Kim-Nilles-Peloso (KNP) alignment \cite{Kim:2004rp} or
  its generalization to models with more than two axions (clockwork
  models~\cite{Choi:2015fiu, Kaplan:2015fuy}). 
\end{itemize}
In the QCD axion context, the three mechanisms and their qualitative
enhancement factors have been studied in Ref.~\cite{Agrawal:2017cmd}.
It is important to keep in mind that all the mechanisms are subject to
different theoretical constraints and may {\em not} generate arbitrarily
large couplings. In particular, we will argue that
\begin{align}
{\rm Large~charges} & \Rightarrow k \lesssim \alpha^{-1}, \nonumber \\
{\rm Clockwork} & \Rightarrow j \lesssim f_a / \mu.
\end{align}
The kinetic mixing case potentially evades the bound on $k$, but this requires a lighter axion field to be present in the EFT. Together, these considerations can significantly constrain the available options for UV completing an effective field theory of axions coupled to gauge fields.

The models we study involve enhanced couplings of an axion-like
particle to the gauge field, which requires the presence of fermions
charged under the gauge group close to the scale suppressing the
operator. The cosmology of these models additionally involves the
production of large gauge fields. An interesting possible consequence
is that these large gauge fields lead to non-perturbative
instabilities via Schwinger pair-production of charged fermions
or non-Abelian gauge fields
~\cite{Kobayashi:2014zza,Hayashinaka:2016qqn,
Hayashinaka:2016dnt,Tangarife:2017vnd,Tangarife:2017rgl,
Hayashinaka:2018amz, Lozanov:2018kpk}. Clockwork theories of multiple axions also have interesting networks of topological defects that can affect the cosmology \cite{Long:2018nsl}.
A detailed study of these phenomena is beyond the scope of this paper.

In Section \ref{sec:mechanisms} we summarize the enhancement with
mechanisms that rely on matter with large charges or kinetic mixing,
and the associated theoretical constraints.  The mechanism that could
lead to the most significant enhancement is the clockwork mechanism.
We will discuss the constraints on it and alignment model in detail in
Section \ref{sec:alignmentandclockwork}.  Section \ref{sec:anbersorbo} discusses large couplings in the Anber-Sorbo
inflation model, which could be clockworked. Section \ref{sec:chromonatural} discusses the
chromonatural inflation model, in which the necessary enhancement for
the axion-gauge field coupling is so large that even the clockwork
mechanism (which is the most efficient way to generate an
exponentially large coupling) could not work unless the inflaton field
range is super-Planckian. We offer concluding remarks in Section \ref{sec:conclusions}.

\section{Mechanisms to Generate Large Axion Couplings}
\label{sec:mechanisms}

In this section, we will discuss issues generating large axion
couplings to gauge fields via a large value of $k$, which may be
obtained by integrating out matter with large charges or through
kinetic mixing.

\subsection{Vector-like Matter with Large Charges}

Let us consider the following interaction
\begin{align}
\frac{k\alpha}{8\pi F_a} a F \widetilde{F} 
\end{align}
Here, $k$ is an anomaly coefficient that depends on the gauge and PQ
charges (and the number of
flavors) of fermions integrated out at the scale $F_a$. 
As discussed in Appendix C of Ref.~\cite{Agrawal:2017cmd}, larger PQ
charges $q_{\rm PQ}$ require small fermion masses $\propto (F_a/\Lambda)^{q_{\rm PQ}}$, 
and it is not possible to
get a very large enhancement by increasing PQ charges alone. We will
not pursue the large PQ charge case further and assume the PQ charge
is 1.

We briefly recall the problem with large gauge charges. We imagine
that the axion interaction is generated by $N_f$ fermions $Q$ as in
the KSVZ axion model \cite{Kim:1979if, Shifman:1979if}:
\begin{align}
  \mathcal{L}
  &=
  \frac{N_f I_2(Q)\alpha}{4\pi F_a} \phi F_{\mu\nu}^a \widetilde{F}^{a,\mu\nu}
\end{align} 
where $I_2(Q)$ is the Dynkin index of the representation. However,
requiring that the gauge theory is perturbative at the scale of the
PQ breaking requires
\begin{align}
  N_f I_2(Q) \alpha \lesssim 1.
\end{align}
Therefore, $k \alpha \lesssim 1$.

In these models the fermions get a mass from the spontaneous breaking
of the PQ symmetry, so that $m_f \lesssim F_a$.

\subsection{Kinetic Mixing}
Kinetic mixing of two axions can
contribute to enhanced couplings to a gauge group \cite{Bachlechner:2014hsa, Shiu:2015uva, Agrawal:2017eqm}. To review
the basic mechanism, consider the following simplified Lagrangian
where the cosmologically evolving axion $a$ kinetically mixes
with a lighter axion $b$,
\begin{align}
  \mathcal{L}
  &=
  \frac12(\partial a)^2 + \frac12(\partial b)^2 
  + \epsilon (\partial_\mu a)(\partial^\mu b)
  + \mu^4 \cos \frac{a}{F_a}
  +\frac{\alpha}{8\pi F_b} b F_a^{\mu\nu} \widetilde{F}^{a,\mu\nu}.
\end{align}
The axion $a$ inherits the coupling
\begin{align}
  \frac{\epsilon F_a}{F_b} 
  \frac{\alpha}{8\pi F_a} a F_a^{\mu\nu} \widetilde{F}^{a,\mu\nu},
\end{align}
so that for $\epsilon F_a / F_b = k \gg 1$, the couplings of $a$
can be enhanced. 

However, we see that there is a coupling of axion $b$ to the
gauge
bosons with a very small value of $F_b$. Therefore, the price for this
enhancement are states
charged under the gauge group which are much lighter than $F_a$. 
The presence of such very light charged fermions can significantly
alter the phenomenology, e.g.~due to Schwinger pair production.

\subsection{Non-compact Fields}
While most of the discussions in the paper will focus
on the case of a compact axion field, the issue of generating a large
coupling coupling can be studied more generally even beyond theories
of compact axions. 
For instance, let us consider the following theory without a
fundamental shift symmetry, 
\be
-\frac{1}{2} m^2 \phi^2 - \frac{c_4}{4!} 
\frac{m^2}{f^2} \phi^4 + \frac{k \alpha}{f} \phi F \widetilde{F} 
+ \cdots,
\ee
with $f$ being the field range of $\phi$.
The dimensionless coefficient $c_4$ is order-one in general. Now,
suppose that we generate the $\phi F \widetilde{F}$ coupling by
integrating out a Dirac fermion beginning with the renormalizable
Lagrangian
\be
m_Q \overline{Q} Q + \iu y \phi \overline{Q} \gamma^5 Q.
\ee
In this case, integrating out the fermion $Q$ will generate an
effective coupling of order
\be
\frac{g^2 I_2(Q)}{16\pi^2} 
\arg(m_Q + \iu y \phi) F \widetilde{F} 
\lesssim \frac{g^2 I_2(Q)}{16\pi^2} \frac{\phi}{f} F \widetilde{F},
\ee
with $I_2(Q)$ the Dynkin index of the representation of $Q$ under the
gauge group. In the second step we have assumed that $m_Q \gtrsim y
\phi \sim y f$, since the value of $\phi$ contributes to the mass of
$Q$. Perturbativity of the theory requires that $\alpha I_2(Q) \lesssim
1$. As a result, we see that again it is difficult to explain the
large $\phi F \widetilde{F}$ coupling relative to $1/f$ (with $f$ the
field range) without a UV completion that contains more structure.
This motivates thinking about mechanisms like alignment or clockwork,
which we will discuss more in detail below, even though we have not
assumed that $\phi$ is a fundamentally compact field. In particular,
here $f$ was merely set by the range of values $\phi$ traverses, not
by its period.

\section{Theoretical Constraints on Alignment and Clockwork Mechanisms}
\label{sec:alignmentandclockwork}
In this section, we discuss a simple theoretical constraint on using the alignment or the clockwork mechanism to generate a large axion coupling to gauge fields. The constraint is, in terms of parameters in the effective Lagrangian of the axion given by Eq.~\ref{eq:axion}
\be
\mu \lesssim F_a. 
\label{eq:constraint}
\ee
We want to emphasize the right hand side is the fundamental period of
the axion, instead of the effective field range, which could be
significantly larger in monodromy models including alignment and
clockwork models. 
The constraint is rooted in the unitarity constraint on a single axion, which will be described in Sec.~\ref{eq:constraintonsingleaxion}. Then we will prove it in the two-axion alignment and multi-axion clockwork model.

\subsection{Unitarity Constraint on Single Axion Model}
\label{eq:constraintonsingleaxion}

Our argument relies fundamentally on the following claim: in any effective theory containing the potential for the single axion
\be
V(a) = \mu^4 \left[1 - \cos\left(\frac{a}{f_a}\right)\right],  \label{eq:cosine}
\ee
we must have $\mu \lesssim f_a$. This follows simply from perturbative unitarity. Expanding around the minimum of the potential, we have
\be
V(a) = \frac{1}{2} \frac{\mu^4}{f_a^2} a^2 - \frac{1}{4!} \frac{\mu^4}{f_a^4} a^4 + \frac{1}{6!} \frac{\mu^4}{f_a^6} a^6 + \cdots,
\ee
so the low-energy amplitude for $a a \to a a$ scattering behaves as 
\be
{\cal M} \approx \left(\frac{\mu}{f_a}\right)^4 \left(1 - \frac{c}{16\pi^2} \frac{\Lambda_{\rm UV}^2}{f_a^2} + \cdots\right),
\ee
with $\Lambda_{\rm UV}$ an ultraviolet cutoff on loops computed with the effective Lagrangian and $c$ some scheme-dependent order-one constant. The first term is the tree-level amplitude, the second comes from beginning with a six-point vertex and sewing up two lines to form a 1-loop diagram, and a series of additional terms appearing with further powers of $\Lambda_{\rm UV}/f$ will arise from higher-point couplings. 

Unitarity requires that the amplitude $\cal M$ be bounded: specifically, the $\ell = 0$ partial wave amplitude $a_0$ should satisfy $|{\rm Re}\, a_0| < \frac{1}{2}$. In perturbation theory, $a_0 \approx -\frac{\cal M}{16\pi}$. Thus, in order for perturbation theory to be approximately reliable we must require
\begin{align}
\left(\frac{\mu}{f_a}\right)^4 &\lesssim 8\pi \quad \Rightarrow \quad \mu \lesssim 2 f_a, \\
\Lambda_{\rm UV} &\lesssim 4\pi f_a.
\end{align}
This already places some constraints on scenarios discussed in the literature; for example, the gauge-flation trajectory of \cite{Adshead:2012qe} is discussed for a parameter point with $\mu = 4 f_a = 4 \times 10^{-2} M_{\rm pl}$. In this case, treating the theory using the classical equations of motion in a cosine potential is not expected to be valid, because unitarity requires large corrections to answers computed perturbatively from the Lagrangian.

One could ask how the constraint $\mu < f_a$ is enforced in UV completions; for example, we can write a renormalizable theory breaking Peccei-Quinn symmetry at some scale $f_a$ and with confinement at some other scale $\Lambda$, and a priori these appear to be independently adjustable parameters. However, if one tries to build a model with $\mu \gg f_a$, one always finds that confinement has important effects on the PQ-breaking dynamics that ensure the bound is respected. Simple examples illustrating this point are discussed in Appendix \ref{app:confinementchecks}.

Let us pause here to discuss a small technical subtlety that in the end has no impact on our arguments. Depending on how the cosine potential \eqref{eq:cosine} is generated, there may already be a monodromy present. For instance, confinement in an $SU(n)$ Yang-Mills theory is expected to generate a potential with $n$ branches of the form $n^2 E(\theta/n)$, where $E$ is a function of period $2\pi n$ despite the fact that $\theta$ has fundamental period $2\pi$ \cite{Witten:1980sp, Witten:1998uka}. The axion potential generated by confinement can be taken schematically to be of the form \eqref{eq:cosine} with $\mu^4 \sim n^2 \Lambda_{\rm conf}^4$ and $f_a = n F_a$, with $F_a$ the fundamental axion period. However, we are not interested in theories with unexplained, parametrically large $n$ (requiring a large number of fundamental fields and implying a UV cutoff $\lesssim M_{\rm pl}/n$). We will focus on small-$n$ gauge groups and the possibility of generating monodromy through products of smaller integers. As a result, we will suppress the factor of $n$ and take the confinement potential to be $\mu^4 \cos(a/F)$, rather than $\mu^4 \cos(a/(nF))$, below. Restoring factors of $n$ will change our final conclusion only by an order-one factor provided all the $n$'s are themselves order-one. In particular, the main argument will rely on the statement that a potential of the form
\be
\mu_{i+1}^4 \cos\left[\frac{1}{n_{i+1}} \left(\frac{N_i a_i}{F_i} + \frac{a_{i+1}}{F_{i+1}}\right)\right]
\ee
serves to set $a_{i+1} = -N_i a_i \frac{F_{i+1}}{F_i}$, a fact which is independent of the $n_{i+1}$ factor.

\subsection{Two-site Alignment Model} 
It is easiest to demonstrate the crucial constraint in Eq.~\ref{eq:constraint} using the simple two-site alignment model that could enhance the axion coupling. Subsequent to the original work of Kim, Nilles, and Peloso \cite{Kim:2004rp}, such models have been studied extensively in e.g.~\cite{Choi:2014rja, Kappl:2014lra, Bai:2014coa, delaFuente:2014aca}. The model is given by 
\be
{\cal L} = \frac{\alpha_{s;1}}{8\pi F_1} a_1 G_1 \widetilde{G}_1 + \frac{\alpha_{s;2}}{8\pi } \left(\frac{Na_1}{F_1} +\frac{a_2}{F_2} \right) G_2 \widetilde{G}_2 + \frac{\alpha}{8\pi F_2} a_2 F \widetilde{F},
\ee
where $G_1, G_2$ are the field strengths of two heavy confining gauge groups. $N > 1$ is an integer and is usually some group theoretical factor in a full model (one full model based on KSVZ fermions could be found in Ref.~\cite{Agrawal:2017cmd}). 
$F$ is the field strength of a gauged $U(1)$ (which is not necessarily our electromagnetic $U(1)$). It is straightforward to generalize the discussion for $U(1)$ to that of a non-Abelian gauge field. $\alpha_{s;1} = g_{s;1}^2/(4\pi), \alpha_{s;2}=g_{s;2}^2/(4\pi), \alpha=g^2/(4\pi)$ are the coupling strengths of the two confining gauge groups and $U(1)$ respectively with $g_{s;1}, g_{s;2}, g$ corresponding gauge couplings.
Below the confinement scales, the effective potential of the two axions is 
\be
V(a_1, a_2) = \mu_1^4 \cos \left(\frac{a_1}{F_1}\right) + \mu_2^4 \cos \left(\frac{Na_1}{F_1}+\frac{a_2}{F_2}\right).
\label{eq:twoaxionpt}
\ee
We need
\be
\mu_1 < \mu_2 \lesssim F_2.
\label{eq:inequalities}
\ee
The first inequality between the two confinement scales $\mu_1$ and $\mu_2$ is needed for enhancing the light axion coupling to photons. If $\mu_1 < \mu_2$, the heavy axion is 
\be
\frac{Na_1}{F_1}+\frac{a_2}{F_2}, 
\ee
up to a normalization factor. Integrating out the heavy axion by setting the heavy axion field to be zero, we have the effective Lagrangian for the light axion, which is mostly $a_2$:
\be
{\cal L}(a_2) =  \mu_1^4 \cos \left(\frac{a_2}{N F_2}\right) + \frac{\alpha}{8\pi F_2} a_2 F \widetilde{F}. 
\ee
Mapping onto Eq.~\ref{eq:axion}, we have 
\be
\mu = \mu_1, \quad F_a = F_2, \quad j=N, \quad f_a = N F_2, \quad {\rm and} \quad k=1.
\ee
To get an enhanced axion-gauge field coupling, Eq.~\ref{eq:enhanced} tells us that $N\alpha \gg 1$. 

The second inequality between $\mu_2$ and $F_2$ is due to the fact that otherwise, the axion quartic coupling would be $\gg 1$ and violates perturbative unitarity, as discussed in the previous section. Another way to understand it is that if $\mu_2 > F_2$, then confinement would break the PQ symmetry first, as discussed in examples in Appendix \ref{app:confinementchecks}. The axion potential is then modified and $F_2$ in Eq.~\ref{eq:twoaxionpt} should be replaced by $\mu_2$. These two inequalities are combined to give us the constraint in Eq.~\ref{eq:constraint}, which in this case is simply $\mu_1 < F_2$. 

In this example, we can derive a separate and even stronger bound: $\mu_1 < \mu_2 \lesssim F_1/N$, again following from unitarity using the second cosine term. If $F_1 \sim F_2$, this is a much stronger bound. In fact, this bound is weaker only when $F_1 > N F_2$ and we have already introduced a large hierarchy of decay constants at the outset, so that alignment is not helping. However, in the clockwork case below the integer $N$ will be a smaller number and there is little difference between the two bounds.

\subsection {N-site Clockwork}
\label{sec:clockwork}
The argument for two-axion alignment model could be generalized to $N$-site clockwork model. This could be mostly easily checked in the confinement tower model~\cite{Agrawal:2017cmd} (a supersymmetric version of which appeared earlier in \cite{Choi:2015fiu}; also see \cite{Coy:2017yex}). Consider $n$ confining gauge groups $SU(n_i)$ with field strengths labeled by $G_i, i= 1,2, \cdots n$ with a Lagrangian
\be
{\cal L} = \sum_{i=1}^{n-1} \frac{\alpha_{s;{i+1}}}{8\pi } \left(\frac{N_i a_i}{F_i} +\frac{a_{i+1}}{F_{i+1}} \right) G_{i+1} \widetilde{G}_{i+1} + \frac{\alpha_{s;1}}{8\pi F_1} a_1 G_1 \widetilde{G}_1 + \frac{\alpha}{8\pi F_n} a_n F \widetilde{F},
\ee
where $N_i \geq 1$'s are integers. 
The potential of the $n$ axions is given by 
\be
V(a_i) = \sum_{i=1}^{n-1} \mu_{i+1}^4\cos\left(\frac{N_i a_i}{F_i} +\frac{a_{i+1}}{F_{i+1}}\right) + \mu_1^4 \cos\left(\frac{a_1}{F_1}\right),
\label{eq:clockwork}
\ee
Note that this potential holds in general clockwork models such as the original model based on a number of scalars with particular quartic couplings.\footnote{In the original model, there are $n$ complex scalars, $\phi_i, i=1,2,\cdots n$ with quartic couplings $\phi_i^3\phi_{i+1}^\dagger + {\rm h.c.}$ Every scalar obtains a vacuum expectation value $\langle \phi_i \rangle$. Their angular modes become the axions with a potential in Eq.~\ref{eq:clockwork}, in which all the $N_i = 3$ and $F_i = \langle \phi_i \rangle$.}

To enhance the couplings of the lightest axion (which is mostly $a_n$) to photons, we need
\be
\mu_1 < \mu_{2,3,\cdots, n}, \mu_n < F_n. 
\label{eq:inequalities2} 
\ee
After integrating out the heavy axions by setting
\be
\frac{N_i a_i}{F_i} +\frac{a_{i+1}}{F_{i+1}} =0, \quad i=1, 2, \cdots n-1,
\ee
we have the effective Lagrangian of the lightest axion $a_n$, 
\be
{\cal L}_{\rm eff} =\frac{\alpha_{s;1}}{8\pi \left(\prod_{i=1}^{n-1} N_i\right) F_n} a_n H_1 \widetilde{H}_1 + \frac{\alpha }{8\pi F_n} a_n F \widetilde{F},
\ee
Mapping onto Eq.~\ref{eq:axion}, we have 
\be
\mu = \mu_1, \quad F_a= F_n, \quad j = \prod_{i=1}^{n-1} N_i, \quad f_a =  \left(\prod_{i=1}^{n-1} N_i\right) F_n, \quad {\rm and} \quad k=1.
\ee 
Again the inequalities in Eq.~\ref{eq:inequalities2} are reduced to Eq.~\ref{eq:constraint}. 

In this case we can derive slightly stronger bounds like $\mu < F_a / N_{n-1}$, but if we take the $N_{i}$ to be order-one numbers this is only a modestly more stringent constraint.

Before moving on to our main examples, let us briefly remark on the original relaxion model \cite{Graham:2015cka}. In this case the potential is of the form $g M^2 \phi + g^2 \phi^2 + \cdots$, where $g$ is a tiny mass scale reflecting small breaking of the $\phi$ shift symmetry and $M$ is a UV cutoff. This is an expansion in $\phi/f_{\rm large}$ where $f_{\rm large} \sim M^2 / g$. From this we read off that the analogue of $\mu^4$ is $g M^2 f_{\rm large} \sim M^4$. If we try to complete the model with clockwork to explain why $f_{\rm large} \gg F$, with $F$ the scale appearing in the $\Lambda(h)^4 \cos(\phi/F)$ term, then Eq.~\eqref{eq:constraint} becomes $M < F$. This constraint is respected in the models of \cite{Graham:2015cka}. We have not explored to what extent this remains true in the large literature of variations on the model.

\section{A Clockworkable Example: Anber-Sorbo Inflation}
\label{sec:anbersorbo}
In some theories, the constraint in Eq.~\ref{eq:constraint} will be a mild constraint. In others, it is a much more difficult constraint to satisfy. An example of the latter case is chromonatural inflation. In this section, we will go through a clockworkable example: the Anber-Sorbo inflation model~\cite{Anber:2009ua}. 

The Anber-Sorbo model is based on natural inflation with an axion-like particle being the inflaton~\cite{Freese:1990rb}. The usual natural inflation model needs the axion field range to be super-Planckian to satisfy the slow-roll condition. What is new in the Anber-Sorbo model is that a large axion-gauge field coupling leads to production of gauge bosons when the axion rolls down the potential. The electromagnetic dissipation slows down the rolling and relaxes the slow-roll constraint. As a result, the field range of the axion inflaton could be sub-Planckian. Below we review the basics of the model and its predictions briefly.

The Lagrangian of the model is
 \be
{\cal L} = \frac{1}{2} (\partial a)^2 + \sum_{i=1}^{{\cal N}} \frac{1}{4} F_{\mu\nu}^iF^{\mu\nu; i} + \mu^4\cos\left(\frac{a}{f_a}\right)+ \sum_{i=1}^{{\cal N}} \frac{k j\alpha}{8 \pi} \frac{a}{f_a} F_{\mu \nu}^i \tF^{\mu \nu;i},
\label{eq:Anber}
\ee
where the potential and interaction terms are the same as in
Eq.~\ref{eq:axion} by expressing $F_a = f_a/j$. 
 The index $i = 1,2, \cdots {\cal
 N}$ labels the 
${\cal N}$ $U(1)$ gauge fields introduced in the model. 
The large number of gauge fields is necessary to get the right
amplitude of the two-point function, as we discuss below. Again
$\alpha=e^2/(4\pi)$ here is the fine structure constant of the gauged
$U(1)$'s (note that in the original paper, $\alpha$ is used as the
overall coupling equivalent to $kj\alpha/(2\pi)$ here). We assume that
the gauge couplings of all the $U(1)$'s are the same without loss of
generality. We also assume the couplings of axion to all the gauge
fields are the same as well. The key parameter that controls the
tachyonic production of gauge fields is $\xi \equiv  \beta
\frac{\dot{a}}{2f_aH}$. 
In the slow-roll solution, it satisfies
\be
\xi  \approx \frac{2}{\pi} \log\left(\xi \frac{14M_{\rm pl}}{({\cal N}\beta)^{1/4}\mu} \right), \quad {\rm with} \quad \beta = \frac{kj\alpha}{2\pi},
\label{eq:xi}
\ee
where ${\cal O}(1) \lesssim \xi \lesssim 20$ and the reduced Planck
scale $M_{\rm pl} = 2.4 \times 10^{18}$ GeV.
The total number of e-folds is given by
\be
N_e = \frac{\beta}{2\xi} \frac{\Delta a}{f_a}  \lesssim \frac{\pi\beta}{2\xi},
\ee
where $\Delta a$ is the range of variation of $a$ during inflation, which is bounded to be $\lesssim \pi f_a$. 

The amplitude of the two-point power spectrum is given by~\cite{Anber:2012du}
\be
\Delta_{\cal R}^2 \approx \frac{0.05}{{\cal N} \xi^2},
\ee
and should match the observation $2.2 \times 10^{-9}$~\cite{Ade:2015lrj}. Since $\xi$ is at most around 20, ${\cal N} > 5.6 \times 10^4$. 
In this model, the generation of photons during inflation sources chiral gravitational waves. As a result, it adds an additional contribution to the tensor to scalar ratio on top of the usual quantum fluctuations of gravitons~\cite{Anber:2012du}
\begin{align}
r &=\frac{1}{P_\xi} \frac{2 V}{3 \pi^2 M_{\rm pl}^4} + 2.7 \times 10^2 \frac{\xi^4}{\beta^2} \left(\frac{V^\prime f_a}{V}\right)^2, \nonumber\\
&= 5.2 \times 10^4 \frac{\xi^6}{\beta} \E^{-2\pi \xi} + 2.7 \times 10^2 \frac{\xi^4}{\beta^2},
\end{align}
where to get the second line, we approximate $V \approx \mu^4, V^\prime f \approx V$ and express $\mu$ in terms of $\xi$ and $\beta$ using Eq.~\ref{eq:xi}. 

To get enough e-folds with $N_e \approx (50 -60)$ and to satisfy the observational constraint $r \lesssim 0.11$~\cite{Ade:2015lrj}, we need a large enhancement of all the axion-photon couplings:
\begin{align}
&{\rm When} \; \xi = 10, \quad \beta \gtrsim 5 \times 10^3 \Rightarrow kj \gtrsim 5 \times 10^4  \left( \frac{0.1}{\alpha}\right) \nonumber \\
&{\rm When} \; \xi = 20, \quad \beta \gtrsim 2 \times 10^4 \Rightarrow kj \gtrsim 2 \times 10^5  \left( \frac{0.1}{\alpha}\right)
\end{align}
To obtain this large enhancement, we need to invoke the clockwork mechanism as described in Sec.~\ref{sec:clockwork}. For instance, it could be realized in the confinement tower model with 10 confining $SU(3)$'s and 10 axion-like particles (the lightest axion being the inflaton) when $\xi = 10$. In this case, the fundamental period of the axion inflaton is 
\be
F_a = \frac{f_a}{j} = 2 \times 10^{-5} f_a  \left( \frac{\alpha}{0.1}\right),
\ee
where we take $k=1$ without loss of generality. The confinement scale $\mu$ is given by, 
\be
\mu = 14 M_{\rm pl} \frac{\xi}{({\cal N} \beta)^{1/4}}\E^{-\pi \xi/2}.
\ee
Since it is exponentially sensitive to the order one parameter $\xi$, the constraint in Eq.~\ref{eq:constraint} and sub-Planckian field range $f_a < M_{\rm pl}$ could be satisfied simultaneously, for instance, when $\xi = 10, \alpha = 0.1, \beta = 5 \times 10^3$ and $f_a = 10^{18}$ GeV, $\mu = 2.8 \times 10^{11}$ GeV and $F_a = 2 \times 10^{13}$ GeV.

In short, the Anber-Sorbo model needs two types of large numbers: {\it a}) ${\cal O}(10^5)$ $U(1)$ gauge fields; {\it b}) each axion-gauge field coupling enhanced by at least ${\cal O}(10^5)$ which could result from clockworking of ${\cal O}(10)$ axions. In addition to predicting chiral gravitational waves, Anber-Sorbo model also predicts a non-Gaussianity $f_{\rm NL}^{\rm equil} = -1.3 \xi$, which is in the range $(5 - 30)$.\footnote{Possible large non-Gaussianity in inflation models with axion coupling to gauge fields have been studied in Refs.~\cite{Barnaby:2010vf, Barnaby:2011vw}. Yet those analyses assume the back-reaction of the gauge field on the inflaton is negligible and inflation potential is flat. Thus their derived constraint on the axion-gauge field coupling does not apply to the Anber-Sorbo model.} This is consistent with current Planck constraint~\cite{Ade:2015ava} and could be confirmed or falsified at CMB stage 4 experiment. Finally, Refs~\cite{Ferreira:2017lnd, Ferreira:2017wlx} also consider the possibility that the large number of gauge fields produced during inflation lead to thermalization and formation of a hot plasma. We stick to the original proposal of Anber and Sorbo and leave the possibility of thermalization for future discussion.

\section{Chromonatural Inflation}
\label{sec:chromonatural}
Chromonatural inflation models feature an axion with a large coupling to gauge fields, requiring a very large enhancement factor $jk$---much larger than the $\mathcal{O}(100)$
number usually stated in the literature. The basic ingredients are an axion which has a potential and a coupling to a non-Abelian gauge field:
\begin{align}
  \mathcal{L}
  &=
  -\frac14 F_{\mu\nu}^a F^{a,\mu\nu}
  +\mu^4 \cos\left(\frac{a}{f_a}\right)
  + \frac{k j \alpha }{8\pi f_a} a F_{\mu\nu}^a \widetilde{F}^{a,\mu\nu}
\end{align}
For this simple analysis, we ignore the fact that we need
to add a Higgs for consistency with inflationary phenomenology
\cite{Adshead:2016omu}. We expect this to make order-one differences in allowed parameters, but not to qualitatively change any conclusion.\footnote{An alternative, more qualitatively different, approach is to introduce a separate inflaton field $\phi$, treating both the axion and the gauge field as spectators during inflation; see e.g.~\cite{Dimastrogiovanni:2016fuu, Fujita:2017jwq, McDonough:2018xzh}. The constraints we discuss should play a role in those models as well, but we will not consider them in detail here.} The original literature on chromonatural inflation introduced a parameter $\lambda \equiv k j \alpha/\pi$ and argued that a modestly large $\lambda \sim 100$ was necessary for the mechanism to work. However, their benchmark model invoked a tiny gauge coupling $\alpha \sim 10^{-12}$, and hence (implicitly) a huge $kj \sim 10^{15}$. (This has previously been pointed out, in passing, in \cite{Heidenreich:2017sim}.) We will now show that the clockwork mechanism is not sufficient to explain this large number.

The gauge field is assumed to have a background,
\begin{align}
  F^a_{0i}
  &=
  \partial_t (\psi(t) {\tt a}(t)) \delta_i^a
  ,\quad
  F^a_{ij}
  =
  -g 
  f_{ij}^a
  (\psi(t) {\tt a}(t))^2
\end{align}
where ${\tt a}(t)$ refers to the scale factor. This classical background spontaneously breaks the product of spatial rotations and internal gauge rotations to the diagonal, preserving a notion of isotropy. Such inflationary backgrounds were first explored in the context of gauge-flation \cite{Maleknejad:2011jw}, which resembles chromonatural inflation but relies on higher dimension operators instead of an axion field (and so is outside the scope of our discussion here).
We are interested in solutions where the energy density is dominated
by the axion $a$, and it has a slow roll solution supported by
friction from the gauge field.
A useful quantity to
parametrize the solutions is
\begin{align}
  w \equiv \left( \frac{k j \mu^4}{6 \pi^2 \Mpl^4}\right)^{1/3}.
\end{align}
(This is approximately $1/m_\psi$ in the notation of \cite{Adshead:2013qp,Adshead:2013nka}.) It turns out that for $w > 1$ fluctuations are
unstable~\cite{Dimastrogiovanni:2012ew, Adshead:2013nka} which will drastically change the
phenomenology, so we restrict to $w\lesssim1$.
The equations of motion imply that (approximately),
\begin{align}
  \frac{\psi}{M_{\rm pl}}
  &\approx
  \frac{\mu^2}{g w\Mpl^2}
  \sqrt{2/3}\sin^{1/3}\theta
  \\
  \frac{1}{f_a}\frac{da}{dN}
  &=
  \frac{2\pi}{k j \alpha}
  \left( \frac{w\cos\theta}{\sin^{1/3}\theta}
  + \frac{\sin^{1/3}\theta}{w\cos\theta}
\right)
\end{align}
with $\theta \equiv a/(2f_a)$.  

It follows that the number of e-folds in chromonatural inflation is
given by 
\begin{align}
N_e(a_0)
&=
\frac{k j \alpha}{\pi}
\int^{\pi/2}_{a_0/2f_a} d\theta\, 
\frac{1}{ 
  { \frac{w\cos\theta}{\sin^{1/3}\theta}
  + \frac{\sin^{1/3}\theta}{w\cos\theta}
  }
}
\lesssim
\frac{3}{2\pi} 
\alpha k j w 
\end{align}
where we have used the fact that $w\lesssim 1$ to derive the final
inequality. 
Our constraint Eq.~\ref{eq:constraint} implies that $j\mu < f_a$,
so that
\begin{align}
  \alpha k j w
  &=
  \alpha \left(\frac{k^4 j^4 \mu^4}{6 \pi^2 \Mpl^4 }\right)^{1/3}
  <
  \alpha k \left(\frac{k f_a^4}{6 \pi^2 M_{\rm pl}^4 }\right)^{1/3}
  \lesssim \left(\frac{k}{6\pi^2}\right)^{1/3}
  ,
\end{align}
where in the last step we have assumed a sub-Planckian field range $f_a \lesssim M_{\rm pl}$ as well as the perturbativity bound $\alpha k \lesssim 1$.

Numerically, this requires that $k$ be quite large: 
\begin{align}
k \gtrsim 6\pi^2 \left(\frac{2\pi N_e}{3}\right)^3 \sim 10^8 (N_e/60)^3.
\end{align}
Because the clockwork mechanism can explain a large $j$ but not a large $k$, this immediately implies that clockwork alone is not sufficient for a viable chromonatural inflation cosmology (if we restrict to sub-Planckian field ranges).  At the same time, the constraint $k\alpha \lesssim 1$ implies that a large value of $k$ is also not sufficient by itself: we would never attain $\lambda \gtrsim 1$ by relying solely on $k$. Therefore, we
conclude that we need \emph{both} a clockwork mechanism as well as (very)
large charges in order to realize chromonatural inflation. The large value of $k$ could potentially arise from kinetic mixing with another, lighter axion, but in this case one should check whether the additional light field has any important phenomenological consequences. We will not consider this possibility in more detail here.

Inflationary phenomenology further restricts the allowed parameter space
significantly.
The other constraints on this parameter space come from the size of
the scalar perturbations, the spectral tilt, and the slow-roll
parameter. In the case of chromonatural inflation, these have been
calculated in some detail~\cite{Dimastrogiovanni:2012st,Dimastrogiovanni:2012ew,Adshead:2013qp,Adshead:2013nka,Papageorgiou:2018rfx}. The scaling behavior
of these quantities is \cite{Adshead:2013nka}
\begin{align}
  \epsilon_H \sim \eta_H \sim 1-n_s
  &
  \sim \frac{1}{N_e}
  \sim \frac{2\pi}{3kj\alpha w}   \label{eq:nsminusone} 
  \\
  \Delta_{\cal R}^2 
  \sim
  \frac{\mu^4}{6 \epsilon_H \Mpl^4}
  &\sim
  \frac{\mu^4}{6 \Mpl^4}
  \frac{3}{2\pi} k j \alpha w
  \,.  \label{eq:DRSq}
\end{align}
We emphasize that these relations can get corrected by
$\mathcal{O}(1)$ factors, especially once we consider the Higgsed
chromonatural model~\cite{Adshead:2016omu}, but the overall scaling
behavior should still hold.

From these equations, we can immediately extract that the value of $\mu$ must be large:
\begin{align}
\mu \sim M_{\rm pl} \left(\frac{\Delta_{\cal R}^2}{N_e}\right)^{1/4} \sim 10^{16}~{\rm GeV}. \label{eq:muchromo}
\end{align}
As discussed in \cite{Heidenreich:2017sim}, this may superficially run afoul of the Weak Gravity Conjecture, which motivates an ultraviolet cutoff at or below $g M_{\rm pl}$ \cite{ArkaniHamed:2006dz}. Nonetheless, this UV cutoff may be of a mild form, associated with the existence of a tower of states charged under the gauge theory \cite{Heidenreich:2015nta,Heidenreich:2016aqi,Montero:2016tif,Andriolo:2018lvp}. This leads to a more stringent UV cutoff at or below $g^{1/2} M_{\rm pl}$, with which chromonatural inflation is at best marginally compatible \cite{Heidenreich:2017sim}. However, in the context of clockwork completions \eqref{eq:muchromo} has an immediate, less conjectural, consequence. The fundamental clockwork constraint \eqref{eq:constraint} tells us that $\mu \lesssim f_a / j \lesssim M_{\rm pl}/j$ (where we again assume a sub-Planckian field range), and hence
\begin{align}
j \lesssim \frac{M_{\rm pl}}{\mu} \sim 10^{2}.
\end{align}
This implies that (in the notation of the original references) the benchmark values of $\lambda$ are only marginally compatible with the clockwork mechanism, and compatibility with this bound further requires us to choose $k \sim \alpha^{-1}$ so that the theory is at or near strong coupling. Notice, in particular, that our arguments require that $k$ is several orders of magnitude larger than $j$, so that clockwork alone falls dramatically short of explaining the large coupling.

\begin{figure}[t]
  \centering
  \includegraphics[width=0.85\textwidth]{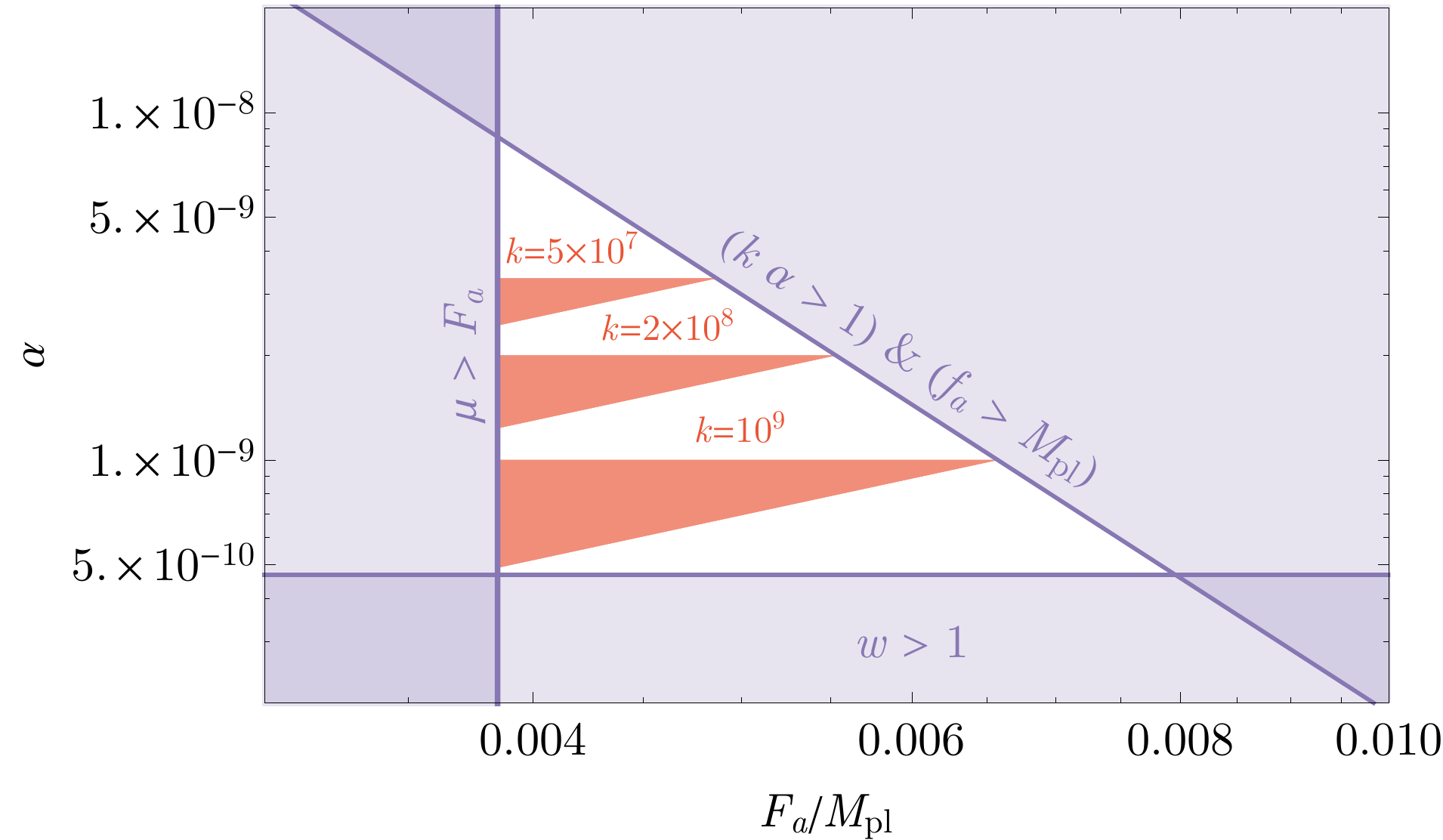}
  \caption{
    Viable parameter space for chromonatural inflation. 
    We assume the number of e-folds of inflation, $N_e = 60$, 
    and
    $\Delta_{\mathcal{R}}^2=2.2\times10^{-9}$ as well as other
    theoretically motivated constraints discussed in the text.
    The red regions show the allowed parameter space for  particular values
    of $k$. The white region is the union of all such regions. The fact that large $k$ is required indicates that clockwork alone cannot explain the large coupling in chromonatural inflation.
  }
  \label{fig:params}
\end{figure}

The two inflationary constraints above can be used to eliminate $\mu$
and $j$, leaving $\{\alpha,F_a, k\}$ as the free
parameters.\footnote{Because we expect that a model compatible with
data in detail will involve various changes to order one factors, as
in \cite{Adshead:2016omu}, we have simply taken the $\sim$ estimates
in \eqref{eq:DRSq} to be equalities and have fixed the number of
e-folds of inflation $N_e = 60$. Our goal
is to convey an order-of-magnitude sense of the constraints, and the
precise order-one factors appearing in our results should not be taken
too seriously.}
There are a number of inequalities that these parameters should
satisfy,
\begin{itemize}
  \item $k \alpha <1$: from perturbativity of the gauge coupling
  \item $w < 1$: for stability of perturbations
  \item $\mu \lesssim F_a$: as derived in equation~\ref{eq:constraint}
  \item $f_a = F_a\, j < \Mpl$: avoiding large field excursions is one
    of the prime motivations of the model
\end{itemize}

These constraints are shown in figure~\ref{fig:params}. (A further constraint, that $j > 1$, is satisfied everywhere in the figure.) For a
particular choice of $k$, the red triangles show the regions allowed
by the constraints above. The upper edge of the triangle corresponds
to
$k \alpha = 1$, whereas the sloping edge corresponds to $f_a = \Mpl$. 
The white region is the union of all such
regions. We see that
the allowed range of parameters is roughly
$\alpha\in\left(10^{-10},10^{-8}\right)$,
$\frac{F_a}{\Mpl}\in (0.004,0.008)$, $j \in (130,260)$ and $k \in \left(10^7,
10^9\right)$.

\section{Conclusions}
\label{sec:conclusions}

Large couplings of axion to gauge fields could lead to very
interesting cosmology. In particular, they may allow alternative
realizations of slow-roll inflation with a steep inflaton potential,
as suggested by Anber-Sorbo and chromonatural inflation models. In
this article, we demonstrate that quite a few axion cosmology models,
including Anber-Sorbo and chromonatural models, actually require
(significantly) larger axion-gauge field couplings than quoted in the
literature, taking into proper account of the presence of the gauge
coupling strength in the axion coupling when gauge fields are
canonically normalized. This observation imposes a highly non-trivial
model-building challenge to axion cosmology: can a large axion
coupling to gauge fields arise in a UV completion without large input
parameters? 

We focus on the two simplest possible methods: inclusion of matter with large charges under the gauge group or PQ symmetry and the clockwork mechanism (equivalently, multi-axion alignment) to significantly extend the effective axion field range beyond its fundamental period. We discuss simple theoretical constraints in either case: in the large charge case, the strong coupling constraint requires the charge times the gauge coupling strength to be around or below one while in the clockwork case, the confinement scale of the non-perturbative dynamics generating the axion potential has to be below the fundamental period of the axion field. These constraints obstruct the UV completion, point towards different feasible parameter space, and may alter the phenomenological predictions.

Finally, we want to comment on possible future directions. In our study, we only use simple parametric approximations for some of the observables. It could be worthwhile to perform a more systematic analysis of the cosmological observables in the possible UV completions. 
Another possible approach to enhance the axion coupling, the kinetic mixing route, is largely unexplored in our paper. It will be interesting to pursue it further and study the dynamics of the additional lighter axion that is needed in this approach.

\acknowledgments{We thank Patrick Draper, David Pinner, and Jure Zupan
  for discussions. PA would like to thank the Kavli Institute for
  Theoretical Physics for their hospitality during the completion of
  this work. PA is supported by the NSF grants PHY-0855591 and
  PHY-1216270.  JF is supported by the DOE grant DE-SC-0010010.  MR is
  supported in part by the DOE grant DE-SC0013607 and the NASA grant
NNX16AI12G. }

\appendix

\section{How Confining Models Ensure $\mu < f_a$}
\label{app:confinementchecks}

We have argued that any effective potential containing the cosine potential \eqref{eq:cosine} requires $\mu \lesssim f_a$. Here we elaborate on this point by showing how the inequality is enforced in some simple examples. If we try to write down a model that will produce $\mu \gtrsim f_a$, we will find that confinement breaks the PQ symmetry and $f_a$ is larger than expected, consistent with this bound. For example, consider the theory where our axion arises from the renormalizable Lagrangian
\be
- \lambda \left(\left|\phi\right|^2 - \frac{1}{2} v_{\rm PQ}^2\right)^2 + \left(y \phi Q \widetilde{Q} + {\rm h.c.}\right),
\ee
with $Q, \widetilde{Q}$ fermions charged in the fundamental and antifundamental representations of an SU($N$) gauge theory. At first glance, the axion decay constant $f_a$ is set by $v_{\rm PQ}$ whereas the scale $\mu$ that will appear in front of the cosine potential is determined by confinement, so we can tune the two parameters independently. However, despite this freedom we cannot attain $\mu \gg f_a$. Confinement produces a quark condensate $\langle Q \widetilde{Q} \rangle \equiv v_{\rm conf}^3$, which is itself a breaking of the PQ symmetry. In turn, this produces an effective tadpole for the scalar field $\phi$, potentially changing its vev. Ignoring radial fluctuations, we can work in terms of two effective pseudoscalar fields $a$ and $\eta'$:
\be
\phi \mapsto \frac{1}{\sqrt{2}} f_a \E^{\iu a(x)/f_a} \quad {\rm and} \quad Q\widetilde{Q} = v_{\rm conf}^3 \E^{\iu \eta'(x)/f_{\eta'}}.
\ee
After confinement the effective potential for these fields is
\be
V(a, \eta') \approx \sqrt{2} y v_{\rm conf}^3 f_a \cos\left(\frac{a}{f_a} + \frac{\eta'}{f_{\eta'}}\right) + \Lambda_{\rm conf}^4 \cos\left(\frac{\eta'}{f_{\eta'}}\right).
\ee
First consider the limit where the tadpole term makes little difference in the vev of $\phi$, i.e.~when $f_a \approx v_{\rm PQ}$. The condition for this to hold is that $y v_{\rm conf}^3 \ll \lambda f^3 \lesssim f^3$. In this case, we have $\mu^4 \sim y v_{\rm conf}^3 f_a \ll f_a^4$, in line with the bound above. On the other hand, it could be that the tadpole term is very significant and that $f_a \gg v_{\rm PQ}$. In this case the vev of $\phi$ is determined primarily by the interplay between the quartic and tadpole terms, and hence
\be
f_a \sim \left(\frac{y}{\lambda}\right)^{1/3} v_{\rm conf}.
\ee
If $\lambda$ is small then this can still be the dominant PQ-breaking vev, but the effective potential for the axion then scales as $y v_{\rm conf}^3 f_a \sim \lambda f_a^4 \lesssim f_a^4$, again in accord with the general bound.

Alternatively, we could consider a theory where the field $\phi$ dominantly has a PQ-stabilizing mass term, i.e.~the Lagrangian is 
\be
-m^2 \left|\phi\right|^2 + \left(y \phi Q \widetilde{Q} + {\rm h.c.}\right),
\ee
and so PQ-breaking is entirely driven by confinement. In that case one has
\be
f_a \sim \frac{y v_{\rm conf}^3}{m^2}.
\ee
To achieve $f_a \ll \mu$ we could attempt to take $m \gg v_{\rm conf}$. However, in that limit, we find that the $\eta'$ becomes lighter than the axion, which has a mass
\be
m_a^2 \sim m^2.
\ee
As a result, we should integrate out the axion and view $\eta'$ as the dynamical field, rather than the alternative.

In this way, we see that concrete models of Peccei-Quinn breaking beginning from renormalizable theories will always respect the perturbative unitarity bounds that we have discussed above.

\bibliography{ref}
\bibliographystyle{jhep}
\end{document}